# Single and multiple doping effects on charge transport in zigzag silicene nanoribbons

Jie Chen,[1] Xue-Feng Wang,[1*] P. Vasilopoulos,[2] An-Bang Chen,[1] and Jian-Chun Wu[3]

[1] College of Physics, Optoelectronics and Energy, Soochow University, 1 Shizi Street, Suzhou 215006, China

[2] Concordia University, Department of Physics, 7141 Sherbrooke Ouest, Montréal, QC, Canada, H4B 1R6

[3] Institute of Nuclear Science and Technology, Sichuan University, Chengdu, 610064, China

* E-mail: xf_wang1969@yahoo.com

A nonequilibrium Green's function technique combined with density functional theory is used to study the spin-dependent electronic band structure and transport properties of zigzag silicene nanoribbons (ZSiNRs) doped with aluminum (Al) or phosphorus (P) atoms. The presence of a single Al or P atom induces quasibound states in ZSiNRs that can be observed as new dips in the electron conductance. The Al atom acts as an acceptor whereas the P atom acts as a donor when it is placed at the center of the ribbon. This behavior is reversed when the dopant is placed on the edges. Accordingly, an acceptor-donor transition is observed in ZSiNRs upon changing the dopant's position. Similar results are obtained when two silicon atoms are replaced by two impurities (Al or P atoms) but the conductance is generally modified due to the impurity-impurity interaction. When the doping breaks the two-fold rotational symmetry about the central line, the transport becomes spin dependent.

## 1. Introduction

A new material, the monolayer honeycomb structure of silicon, called silicene, has been recently synthesized[1-4] on Ag, $ZrB_2$ and Ir surfaces and has attracted considerable attention.[5-8] The reason is that the observed energy band has Dirac cones. Though this band structure is still in question[9], it has been predicted that, similar to graphene, free-standing silicene has Dirac cones with a linear electronic energy dispersion near the Fermi energy[10]. Under an external vertical electric field monolayer graphene remains zero-gap semi-metallic because its two sublattices remain equivalent. In contrast, the most stable silicene has a low-buckled structure resulting from the large ionic radius of silicon. This results in a height difference between the two Si atoms in the primitive cell and leads to a gap under an external electric field since the atoms in a buckled structure are not equivalent. This gap has a different origin than that in bilayer or multilayer graphene under a bias.[11] All that and silicene's compatibility with silicon-based electronic technology have led already to various studies such as the spin-Hall effect,[5] the anomalous Hall effect,[12] the capacitance of an electrically tunable silicene device, [13] etc.

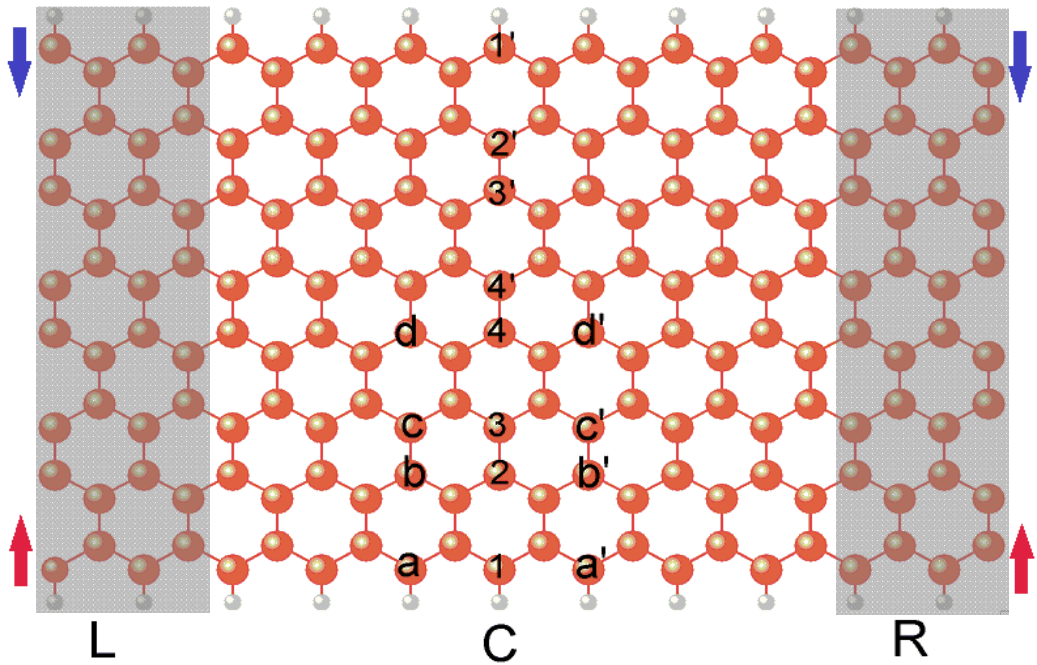

FIG. 1. (Color online). A 8-ZSiNR device consisting of two electrodes (L, R, grey areas) and the scattering region C between them. The big (small) spheres are the Si (H) atoms. The indices 1, 2,…, 1', 2' ,…., a, b,…., and a', b',…., show the doping sites. The down (up) arrows describe the down (up) polarization of the upper (lower) edges.

Graphene nanoribbons have been investigated for several years due to many properties that render them potential materials for nanodevices.[14-17] Recently, for the same reasons silicene nanoribbons (SiNRs) have attracted much attention. The fabrication of SiNRs has been realized[18] and the giant magnetoresistance of zigzag SiNRs (ZSiNRs) has been reported.[19-21] Other studies dealt with the width dependence of the band structure of ZSiNRs[22] and their thermoelectric properties.[23] A very recent study[24] calculated the band structure of doped ZSiNRs and armchair SiNRs (ASiNRs) with single and multiple dopants.

There are still many aspects of silicene that justify further studies. No electrodes were attached to the SiNRs in Refs.[22-24] and electron transport was not considered. In this work we study the spin-dependent band structure and ballistic transport properties of 8-ZSiNRs doped with aluminum (Al) or phosphorus (P) atoms. The 8-ZSiNR is in its antiferromagnetic (AFM) state, the ground state in the absence of an external field, as indicated in Fig. 1 by the blue (spin-down) and red (spin-up) arrows. We consider single and double doping at various sites, as shown in Fig. 1, and evaluate the conductance and in some cases the band structure. We support these results by an evaluation of the corresponding wave function and of the local density of states (LDOS) and compare them with those of undoped ZSiNRs. The presence of a single Al or P atom induces quasibound states in ZSiNRs that can be observed as dips in the electron conductance. An unusual acceptor-donor transition is observed in ZSiNRs.

In Sec. 2 we present the model and in Sec. 3 the results obtained using the non-equilibrium Green's function (NEGF) method combined with density functional theory (DFT). A summary follows in Sec. 4.

2. **Model and method**

The geometric structures of pristine and doped 8-ZSiNRs , consisting of a left and right electrode and the central scattering region between them, are shown in Fig. 1. A vacuum layer thicker than 15 Å is used to eliminate possible mirror interaction and the edge Si atoms are passivated by hydrogen atoms so as to saturate the dangling bonds of Si atoms. We have optimized the structures by employing the DFT in the generalized gradient approximation with the Perdew-Burke-Ernzerhof exchange-correlation functional as implemented in the Atomistix ToolKits (ATK) and the Vienna *ab initio* simulation packages. All structures are fully relaxed until the forces felt by each atom are smaller than 0.02 eV/Å.

The spin-dependent band structures and ballistic transport properties of ZSiNRs are then calculated by DFT combined with NEGF formalism, as implemented in the ATK package. In the computation we use the functionals in the local-density approximation with the Perdew-Zunger parameterization, a double-$\zeta$ plus one polarization orbital basis set, and a 1 × 1 × 500 Monkhorst-Pack k-point mesh. The grid mesh cutoff is set to 250 Ry and the temperature of the electrodes to 300 K. The spin-dependent conductance is evaluated by the Landauer formula[29,30]

$$G_\sigma(E) = \frac{e^2}{h} T_\sigma(E) = \frac{e^2}{h} Tr[\Gamma_L G^R \Gamma_R G^A]_\sigma$$

with $T_\sigma$ the transmission for spin σ, $\Gamma_L$ ($\Gamma_R$) the broadening matrix due to the left (right) electrode, and $G^R$ ($G^A$) the retarded (advanced) Green's function.

## 3. Results and discussion

There are various different spin-polarized states for the same atomic structure of a ZSiNR. In order to determine the ground state, we have calculated the total energies for different magnetic states of *n*-ZSiNRs of width *n*=3, 4, 5, ..., 13 including the nonmagnetic (NM), the ferromagnetic (FM), and the antiferromagnetic (AFM) state. In the FM and AFM states, the two edges of the ZSiNRs are magnetized in the same and opposite directions, respectively. As illustrated in Table 1, the AFM state has the lowest total energy and is the ground state in the absence of an external field, similar to the case of zigzag graphene nanoribbons. In the following we will assume that a AFM 8-ZSiNR is in its ground state.

**Table 1** Energy differences between the NM and AFM states (second column) and between the NM and FM states (third column) for different ribbon widths (first column). These differences are calculated with respect to a primitive unit cell of a 8-ZSiNR.

| Width $n$ | $\Delta E_{NM-AFM}$ (meV/unit) | $\Delta E_{NM-FM}$ (meV/unit) |
|---|---|---|
| 4 | 56.1 | 44.4 |
| 5 | 75.8 | 71.4 |
| 6 | 81.5 | 80.0 |
| 7 | 82.4 | 81.6 |
| 8 | 82.6 | 81.8 |
| 13 | 85.0 | 84.5 |

1. *Single dopant.* Stability and doping type

In Table 2 we present the formation energies to show the stability of the systems and the transferred electrons to show the doping type for different doping sites and elements. The formation energy for substitution is calculated as: $E_{form} = E_{ud} + mE_{Al} + nE_P - E_d - (m+n)E_{Si}$. Here $E_{ud}$ and $E_d$ are the total energies of an undoped and doped ZSiNR, respectively; the $E_{Si}$, $E_{Al}$ and, $E_P$ are the energies of isolated Si, Al and, P atoms, respectively. $m$ and $n$ are the numbers of Al and P atoms in the doped ZSiNR, respectively. The formation energies of a single Al or P substitution at four different sites of 8-ZSiNR are summarized in Table 2. The results show that all optimized structures considered in the manuscript are stable. For the group-III (group-V) doping element Al (P), the formation energy increases (decreases) with the doping site from edge to center. It is interesting to note that a doping type transition occurs in ZSiNRs when the doping site is varied. As shown in Table 2, when an Al atom is doped on the lower edge (Fig. 1, site 1), the Al atom donates 0.45 electrons to the host Si atoms around and works as a donor. In contrast, when an Al atom is doped on the center (Fig. 1, site 4), it accepts about 0.51 electrons from the host Si atoms and acts as an acceptor impurity. The transition from acceptor to donor occurs for P doping atoms.

As discussed above, the Al- (P-) doped ZSiNRs exhibit an acceptor (donor) character and, rather unexpectedly, a donor (acceptor) one when the dopant is placed at the center and on the edges, respectively. A similar effect has been reported for B and N doping in carbon nanotubes[25,26] and in zigzag graphene nanoribbons[16]. It has been expected to be related to the competition between two different phenomena, the Coulomb interaction of charge carriers with the ion impurity and the correlation between charges at the edges.

**Table 2**. Calculated formation energies $E_{form}$ (in eV) and electrons (− e) lost from each impurity atom in the single dopant case.

| Impurity site | Al 1 | Al 2 | Al 3 | Al 4 |
|---|---|---|---|---|
| $E_{form}$ (eV) | 1.54 | 18.41 | 20.92 | 30.82 |
| Lost electrons | 0.4533 | -0.5282 | -0.4961 | -0.5107 |
| Impurity site | P 1 | P 2 | P 3 | P 4 |
| $E_{form}$ (eV) | 10.78 | 4.41 | 3.85 | 0.76 |
| Lost electrons | -0.2456 | 0.2313 | 0.2325 | 0.252 |

2. *Single dopant.* Conductance

Figure 2 shows the conductance G as a function of the energy, measured from the Fermi level, in case only *one* Si atom is replaced by an impurity atom. When the Al atom lies on the lower edge (Fig. 1, site 1), conductance dips in both spin channels can be observed at around 0.3 eV above $E_F$, see Fig. 2(a). *At the top* of the valence band (VB), the conductance of the spin-up channel, compared to that of the undoped ribbon, drops from 2 to 1 whereas that of the spin-down channel remains nearly unchanged. This behavior is reversed *at the bottom* of the conduction band (CB): here the conductance of the spin-up channel is almost unchanged but that of the spin-down channel is rapidly suppressed. When the Al atom is placed in the inner part of the ribbon (sites 2, 3, and 4 in Fig. 1), the conductance dips in both spin channels are found below $E_F$. Comparing to the doping case at site 1 (Fig. 2(a)), we also observe that the conductance close to the top of the VB and to the bottom of the CB is suppressed gradually in both spin channels since the Al atom is located closer to the ribbon center, see Fig. 2(b)-(d). In Conclusion, as the impurity moves from the edges to the center, the conductance for both spin channels is progressively suppressed. This also occurs in zigzag graphene nanoribbons doped with B or N atoms.[17]

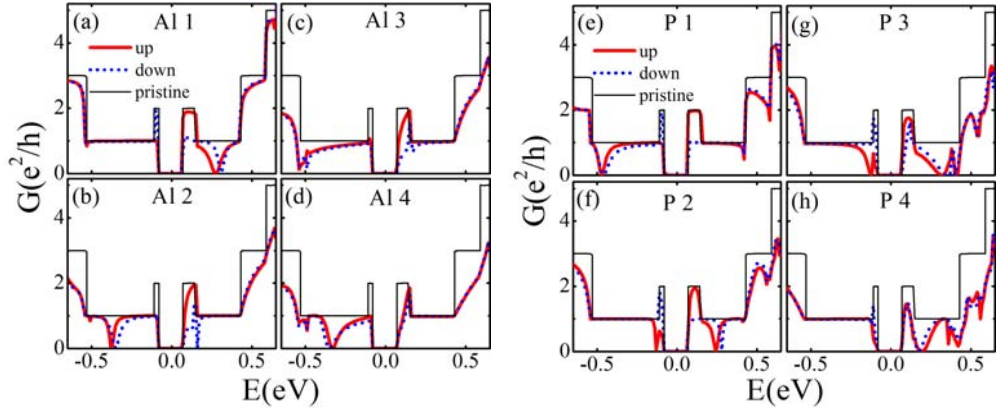

FIG. 2. Left panels. The conductance G of a ZSiNR, in units of $e^2/h$, with Al atoms substituted at different sites (a) 1, (b) 2, (c) 3, and (d) 4. Right panels. As on the left panels with a P atom at sites (e) 1, (f) 2, (g) 3, and (h) 4.

In P doping cases, similar results are obtained. The main difference between the two is that the conductance dips created by P always occur at energies, measured from $E_F$, opposite to those generated by Al, see Figs. 2(e)-(h). For instance, Al doping on site 1 produces a dip above $E_F$ whereas below $E_F$ the conductance remains practically unchanged compared to that of the pristine ribbon. In contrast, the P doping at site 1 produces a dip below $E_F$ and the conductance remains practically unchanged above $E_F$. We noted that in all cases investigated, the breaking of the ribbon mirror symmetry makes the edge states on the two sides of the ribbon different from each other. The spin degeneracy is then broken and the conductance becomes spin dependent.

3. *Single dopant.* Electronic structure

To understand the formation of the impurity states in the doped two-probe systems, we construct bulk systems, with unit cell of the same size as the central scattering region of the two-probe systems, and calculate their band structures and wave functions. Figure 3 shows (a), (c) the band structures and (b), (d) wave functions of a perfect and an edge-doped ZSiNR, respectively, with the Al atom at the center of its lower edge. The four bands near the charge neutrality point (CNP) are labeled 1, 2, 3, 4. Due to the spin degeneracy of the edge states, the energy bands of the pristine ribbon overlap. Comparing the wave functions at the Γ point of the undoped and doped ribbons, we observe that the wave functions corresponding to the bands 1, 3, and 4, for spin-up electrons, are only weakly affected by the impurity, indicating that there is no mixing with neighboring bands. But band 2 is shifted above $E_F$ in the doped ribbon and mixes with neighboring bands; this results in the opening of an energy gap near the energy at which the conductance dips appeared. For spin-down electrons, relative to the perfect case the wave functions corresponding to the bands 1, 2, and 3 are almost unchanged but band 4 mixes with neighboring bands and opens an extra energy gap slightly above the spin-up gap.

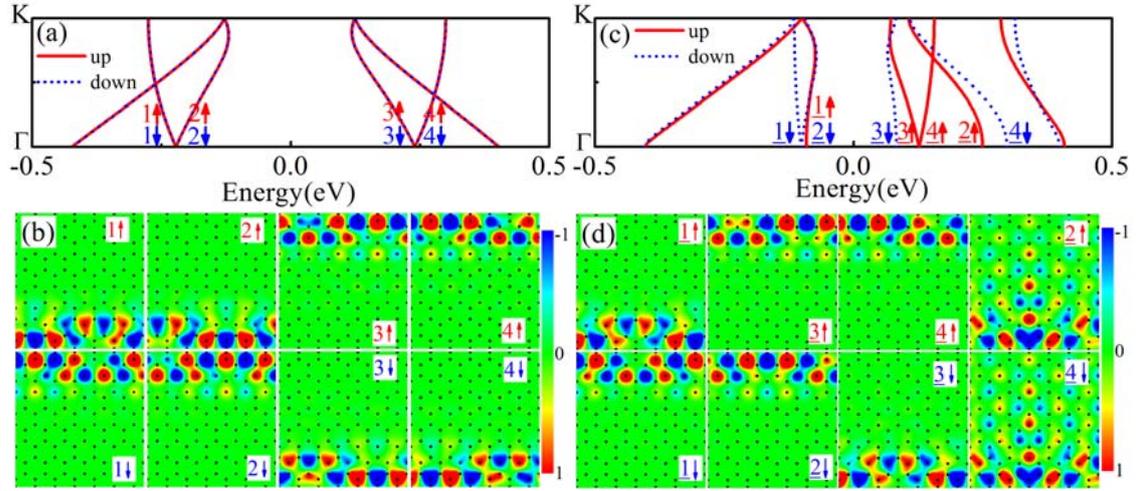

FIG. 3. Band structures of bulk systems with unit cell of the same size as the central region in Fig. 1 for (a) a perfect ZSiNR and (c) a ZSiNR with a single Al dopant at the center of its lower edge. The corresponding wave functions at the Γ point, near the CNP, are shown in (b) and (d).

To further understand the change of the conductance in both spin channels at the top of the VB and at the bottom of the CB, we plot the corresponding spin-dependent local density of states (LDOS) of the two-probe systems, at some especial energy values, in Fig. 4. From Fig. 4(a), (b) we infer that spin-up and spin-down electrons are mainly transported via opposite edge atoms, which agrees with previous reports.[27,28] We also see that at the top of the VB, spin-up electrons are mainly transported through the lower-edge silicon atoms. The corresponding LDOS decreases from the two ends of the scattering region toward the Al doping position; this means that it's difficult for an electron to be transported from one end to another. The presence of the Al dopant breaks the extended nature of the lower edge and results in the reduction of the conductance of the spin-up channel. In contrast, spin-down electrons are transported through the upper edge atoms since the upper edge states are not affected, see Fig. 4(b). Accordingly, the conductance of the spin-down channel at the top of the VB is unchanged. However, at the bottom of the CB the situation is reversed, we see that the conductance of the spin-down channel is suppressed, see Fig. 4(b), with the corresponding states localized on the lower edge, whereas it remains unchanged for the spin-up channel localized at the upper edge, see Fig. 4(c).

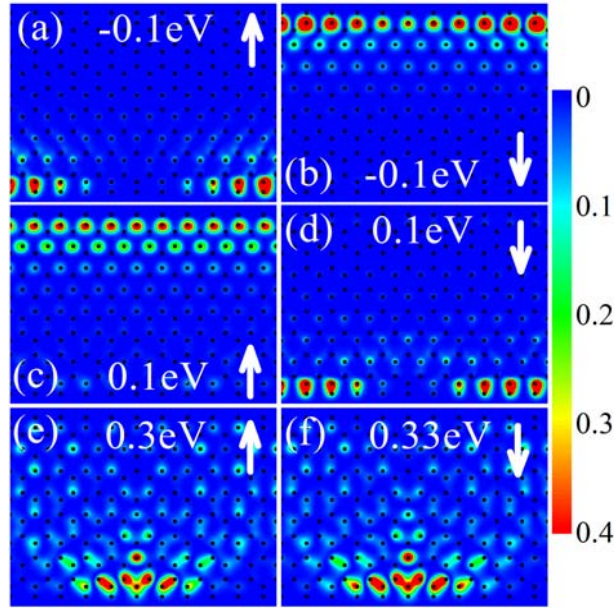

FIG. 4. LDOS of two-probe system at some special energy values for a ribbon edge-doped with Al atom. The arrows indicate the direction of spin polarization.

The LDOS, pertinent to the conductance dips, near E = 0.3eV for both spin channels, is shown in Fig. 4(e)-(f). The corresponding quasibound states are localized around the impurity and decay rapidly away from it towards the ends. These quasi-bound states produce a strong backscattering of the electrons for specific resonant energies and the conductance dips are observed in this energy range, see Fig. 2(a).

### 3. *Double dopant.* Conductance

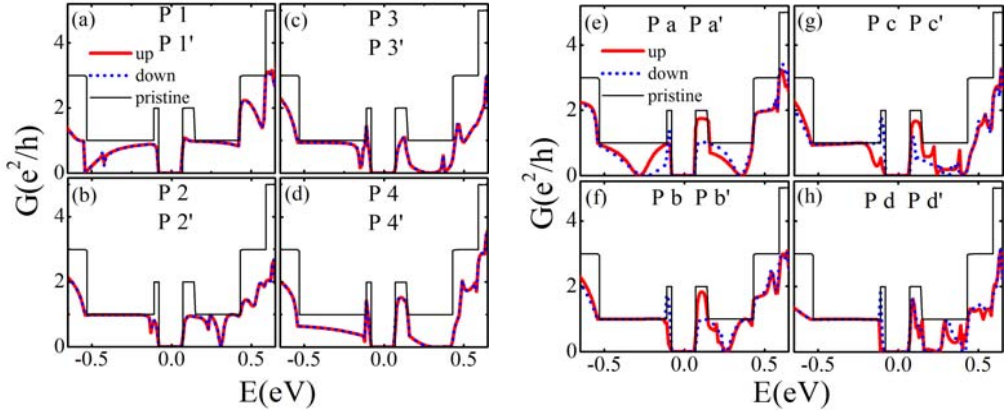

FIG. 5. The conductance G of a ZSiNR with two P atoms substituted at different sites (a) 1 and 1', (b) 2 and 2', (c) 3 and 3'; (d) 4 and 4', (e) a and a', (f) b and b', (g) c and c', (h) d and d'.

In the case of double doping we study systems where two Si atoms in ZSiNRs are replaced by two Al or P atoms, labeled as Al/Al or P/P. As will be shown, the conductance exhibits more features compared to that of the single dopant case considered so far. As an example, consider two P atoms substituted symmetrically in a ZSiNR, at sites 1 and 1' or at 2 and 2',…, see Fig. 1. The conductances are shown in Fig. 5(a)-(d). It can be seen that the spin-up and spin-down channels are both fully degenerate, indicating nonmagnetic properties, different from the single P dopant case shown in Fig. 2. It happens because the two-fold rotational ($C_2$) symmetry of the system in those doping configurations remains intact. This agrees well with the N/N doping result of Ref. [24]. In the second example, we place the two impurities on one edge at equal distances from the edge's center and that of the scattering region, at sites a and a' or at b and b',…, see Fig. 1. In both cases we can still find conductance dips, whose positions are mostly similar to those for a single impurity placed at the same sites, since there is little interaction between the impurities. However, in general the conductance is less regular than that for single doping due to the interference between the two dopants that have similar energies.

We expect that by increasing the number of impurities, the character of the quasibound states will become more complex, due to the enhanced impurity-impurity interaction, and this will be reflected in the conductance. Thus, well-shaped conductance dips or gaps can be achieved only for single impurity doping or when a large difference exists in the energies of the quasibound states associated to the individual dopants. For instance, this is the case with Al/Al or P/P doping at sites 1 and 4, see Fig. 6. As can be seen, the single-dopant conductance features of Fig. 2 are recovered because the energies of the quasibound states of the two Al atoms are well above and below $E_F$, respectively, indicating that the effect of impurity-impurity interaction is negligible.

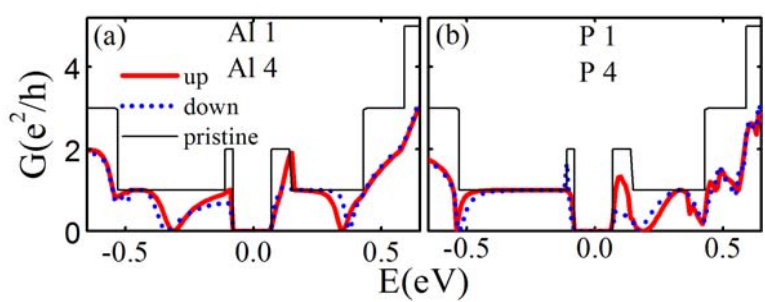

FIG. 6. The conductance of a ZSiNR with two Al impurities doped at sites 1 and 4 shown in (a) and in (b) for two P impurities.

Now, we consider Al and P co-doping in ZSiNRs. Also in this case the impurities are placed at the same sites as in the Al/Al or P/P case. The conductances are shown in Fig. 7. In both cases we observe that the conductance Al and P dips are almost symmetrically located at opposite sides of the Fermi level. Thus, the quasibound states induced by the dopants are almost a superposition of those obtained by single Al and P doping. Interestingly, the dips practically disappear when the Al and P atoms are placed at sites 4 and 4', respectively, see Fig. 7(d). This phenomenon may be attributed to the charge compensation effect between Al and P impurities.

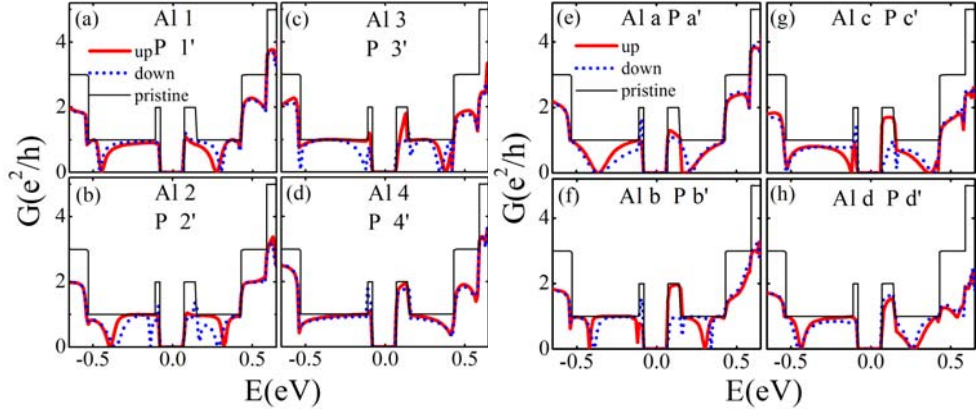

FIG. 7. Conductance versus energy with one Al and one P impurity at different positions: (a) 1 and 1', (b) 2 and 2', (c) 3 and 3', (d) 4 and 4', (e) a and a', (f) b and b', (g) c and c', (h) d and d'.

**4. Summary**

We studied the effects single and double substitutional dopants, Al or P atoms, have on charge transport in zigzag silicene nanoribbons (ZSiNRs) when they are in their antiferromagnetic ground state. We considered various positions of these dopants in the central region of 8-width ZSiNRs and performed first-principle calculations.

We found that Al atoms act as acceptors whereas the P atoms act as donors when they are placed at the center of the ribbon. This behavior is reversed when the dopants are placed on the edges. Thus, an acceptor-donor transition can occur upon changing the dopant's position. New dips in the conductance occur for single dopants when compared to that of undoped 8-ZSiNRs. All single-dopant results were supported by the evaluation of the quasibound states they create and the corresponding local density of states. The features for Al and P dopants are similar except for the difference of opposite doping types.

**Acknowledgments**

This work was supported by the National Natural Science Foundation in China (Grant Nos. 11074182, 91121021, and 11247276) and by the Canadian NSERC Grant No. OGP0121756.